\documentclass{ws-mpla-arXive}

\newcommand{\diag}{{\rm diag}}
\newcommand{\ev}{{\rm eV}}

\newcommand{\gev}{{\rm GeV}}

\begin{document}

\markboth{T. Araki and C. Q. Geng}
{Leptogenesis with Friedberg-Lee Symmetry}

%%%%%%%%%%%%%%%%%%%%% Publisher's Area please ignore %%%%%%%%%%%%%%
%\catchline{}{}{}{}{}
%%%%%%%%%%%%%%%%%%%%%%%%%%%%%%%%%%%%%%%%%%%%%%%%%%%%%%%%%%%%%%%%%%%

\title{Leptogenesis with Friedberg-Lee Symmetry
}

\author{\footnotesize Takeshi Araki
\footnote{Talk given at International Workshop on Dark Matter, Dark Energy
and Matter-Antimatter Asymmetry, Hsinchu, Taiwan, 20-21 Nov. 2009.
}~~~ and C. Q. Geng
}

\address{Department of Physics, National Tsing Hua University, 
Hsinchu, Taiwan 300
}

\maketitle

%\pub{Received (Day Month Year)}{Revised (Day Month Year)}

\begin{abstract}
We consider the $\mu-\tau$ symmetric Friedberg-Lee (FL) symmetry for the 
neutrino sector and show that a specific FL translation leads to the tribimaximal 
mixing pattern of the Maki-Nakagawa-Sakata (MNS) matrix.
We also apply the symmetry to the type-I seesaw framework and address the 
baryon asymmetry of the universe through the leptogenesis mechanism.
We try to establish a relation between the net baryon asymmetry and CP 
phases included in the MNS matrix.

\end{abstract}

%\ccode{PACS Nos.: include PACS Nos.}

\section{Introduction}
From the neutrino oscillation experiments, we currently know for sure that 
the neutrinos have tiny but non-zero masses and mix with each other through 
the Maki-Nakagawa-Sakata (MNS) mixing matrix.
%Especially, 
In particular, the mixing pattern of the MNS matrix is
%have well been 
almost coincident 
with the so-called tribimaximal (TBM) mixing\cite{TB}, given by
\begin{eqnarray}
V_{TB}=\frac{1}{\sqrt{6}}
 \left(\begin{array}{ccc}
  2 & \sqrt{2} & 0 \\
  -1 & \sqrt{2} & -\sqrt{3} \\
  -1 & \sqrt{2} & \sqrt{3}
 \end{array}\right)\label{eq:TB}.
\end{eqnarray}
However, 
%we do not  have 
%confirmed 
there is no definitive theory to generate the neutrino 
masses and TBM mixing yet.
The (type-I) seesaw mechanism is one of the most plausible extensions 
of the  standard model (SM) to produce  tiny neutrino masses while as a bonus 
it can explain the baryon asymmetry of the universe (BAU) through 
the leptogenesis mechanism\cite{LG}.
In the leptogenesis scenario, CP asymmetry is an essential ingredient and 
it is related to the Dirac and Majorana phases in the MNS matrix.
Nevertheless, it is usually very hard to connect them directly because the model 
also includes some high-energy phases\cite{highCP,Xing}, which are associated with 
the heavy right-handed Majorana neutrinos, and the CP asymmetry of the 
leptogenesis depends on the phases as well.
In order to establish a direct relation between the CP asymmetry and phases 
in the MNS matrix, we need to reduce as many complex parameters as 
possible in the model.
In Refs. \refcite{flavorLG1,flavorLG2}, a family symmetry is used to minimize 
the number of arbitrary parameters in the Yukawa sector.
Another possibility along this direction is to consider the so-called 
two-right-handed neutrino (2RHN) seesaw model\cite{32-1,32-2}, in which 
the number of parameters is less numerous than the ordinary seesaw model. 
In addition, spontaneous\cite{MChen} and dynamical\cite{YLin} 
CP violating approaches have been proposed.

In this talk, we introduce the Friedberg-Lee (FL) symmetry\cite{FL,FL2,hFL}
for the neutrino sector and show our results based on Ref. \refcite{FL-LG}.
The FL symmetry is a translational (hidden) family symmetry and 
some detailed analyses for the neutrino sector have been discussed in
Refs. \refcite{gFL,FLxing1,FLxing2,FLxing3} and \refcite{XG}.
Remarkably, as pointed out in Ref. \refcite{jarlskog}, the introduction of 
the FL symmetry to the right-handed neutrinos (RHNs) suggests the existence of 
a non-interacting massless RHN, and the theory comes down to the 2RHN
seesaw model.
This, in fact, motivates us to examine the leptogenesis in the context of the 
FL symmetry.
%since the number of parameters included in the 2RHN model is less 
%numerous than that of the ordinary seesaw model.
We will consider the $\mu-\tau$ symmetric FL symmetry\cite{twFL} 
and show that the TBM mixing can be explained by the specific pattern of 
the FL translation.
In particular, we apply our scheme to the seesaw mechanism and address the BAU via 
the leptogenesis mechanism.

\section{Friedberg-Lee Symmetry and Neutrino Mixing}
We start our discussion with the Lagrangian of the Majorana neutrino mass term
\begin{eqnarray}
-{\cal L}_\nu = \overline{\nu^c}_{i}{\cal M}_{ij}^{\nu}\nu_{j}+h.c.\ ,
\end{eqnarray}
where the subscripts $i$ and $j$ stand for family indices.
Here, we take ${\cal M}^\nu$ as a real matrix and consider 
the diagonal basis of the charged leptons.
In this basis, we impose the FL symmetry on the Majorana neutrinos 
as follows 
\begin{eqnarray}
 \nu_i \rightarrow \nu_i^{'}=\nu_i + (1,\eta,\eta\xi)^T z\ ,\label{eq:FL}
\label{uni}
\end{eqnarray}
where $z$ is a space-time independent Grassmann parameter, $z^2 =0$, and 
$\eta$ and $\xi$ are c-numbers.
%Consequently, t
The mass matrix takes the form
\begin{eqnarray}
{\cal M}^\nu =
 \left(\begin{array}{ccc}
  B\eta^2 + C & -B\eta & -C/(\eta\xi) \\
  -B\eta & A\xi^2 + B & -A\xi \\
  -C/(\eta\xi) & -A\xi & A + C/(\eta\xi)^2
 \end{array}\right).\label{eq:FL-M}
\end{eqnarray}
We note that the mass matrix has one zero-eigenvalue\cite{XG},
% and this fact 
which is ensured 
by the FL symmetry\footnote{
In general, the FL symmetry leads to one zero-eigenvalue because the 
mass matrix needs to satisfy the condition 
$(1,\eta,\eta\xi)_i M_{ij}=0$ to keep the invariance.
In other words, the three-dimensional vector $(1,\eta,\eta\xi)$ corresponds to 
the eigenvector of the zero-eigenvalue.
}.
According to the procedure in Ref. \refcite{FL2}, we have assumed the relation 
$C=B\eta^2\xi^2$.
Consequently, the MNS matrix can be expressed with only $\eta$ and $\xi$:
\begin{eqnarray}
V_{MNS} =
 \left(\begin{array}{ccc}
  \cos\sigma & -\sin\sigma & 0 \\
  \sin\sigma\cos\rho & \cos\sigma\cos\rho & -\sin\rho \\
  \sin\sigma\sin\rho & \cos\sigma\sin\rho & \cos\rho
 \end{array}\right) \label{eq:FL-MNS}
\end{eqnarray}
with $\eta=\tan\sigma\cos\rho$ and $\xi=\tan\rho$.
That is, the pattern of the neutrino mixing is governed by the FL symmetry 
%given 
in Eq. (\ref{eq:FL}).
% and t
Thus, by comparing Eq. (\ref{eq:FL-MNS}) with 
the neutrino oscillation data, we can deduce the values of $\eta$ and $\xi$.
In particular, the TBM mixing pattern corresponds to 
$\eta=-1/2$ and $\xi=1$. 
Interestingly, in this case, the relation assumed above becomes 
$C=B/4$ and it can be realized by imposing the $\mu-\tau$ symmetry.
Namely, the FL symmetry, with $\eta=-1/2$ and $\xi=1$, combined with the 
$\mu-\tau$ symmetry can naturally lead to the TBM mixing.
So, we define the symmetry as follows 
\begin{eqnarray}
\left(\begin{array}{c}
\nu_e \\ \nu_\mu \\ \nu_\tau
\end{array}\right)
\rightarrow\left(\begin{array}{ccc}
1 & 0 & 0 \\
0 & 0 & 1 \\
0 & 1 & 0
\end{array}\right)
\left(\begin{array}{c}
\nu_e \\ \nu_\mu \\ \nu_\tau
\end{array}\right)
+
\left(\begin{array}{c}
-2 \\ 1 \\ 1
\end{array}\right)z\label{eq:twFL}
\end{eqnarray}
and call it twisted FL symmetry\cite{twFL}.
Note that the shift part of Eq. (\ref{eq:twFL}) is multiplied by a factor of $-2$ 
because the overall factor is irrelevant in this discussion.

\section{Twisted Friedberg-Lee Symmetric Seesaw Model}
\subsection{Framework of model}
We apply the twisted FL symmetry to the conventional 
(type-I) seesaw framework with three RHNs.
The relevant Lagrangian is given by
\begin{eqnarray}
 -{\cal L}_{\rm seesaw}=
  Y_D \bar{L}_{L} {\tilde H} \nu_{R}
  +\frac{1}{2}M_R \overline{\nu^c}_R \nu_R+h.c.\ ,
\end{eqnarray}
where we have omitted family indices.
We assume the diagonal charged lepton mass matrix again
and impose the twisted FL symmetry on both the 
right- and left-handed neutrinos.
Due to the symmetry, the Majorana mass matrix takes the form
\begin{eqnarray}
M_R=
 \left(\begin{array}{ccc}
  B/2 & B/2 & B/2 \\
  B/2 & A+B & -A \\
  B/2 & -A & A+B
 \end{array}\right).\label{eq:MR}
\end{eqnarray}
In addition to the twisted FL symmetry, we introduce 
a $Z_2$ symmetry for the lepton doublet and charged singlet of the first family 
in order to reproduce a realistic neutrino mass hierarchy.
As a result, the Dirac mass matrix is given by
\begin{eqnarray}
Y_D=
 \left(\begin{array}{ccc}
  0 & 0 & 0 \\
  0 & \alpha & -\alpha \\
  0 & -\alpha & \alpha
 \end{array}\right)\label{eq:YD}.
\end{eqnarray}
The Majorana mass matrix in Eq. (\ref{eq:MR}) can be diagonalized by the 
TBM matrix in Eq. (\ref{eq:TB}), so that
\begin{eqnarray}
 D_R \equiv (P V_{TB}^T) M_R (V_{TB}P) &=& \diag(M_1, M_2, M_3)\nonumber\\
  &=&\diag(0, 3/2|B|, |2A+B| ),
\end{eqnarray}
where $P=\diag(1,e^{i\phi_R /2},1)$ is a diagonal phase matrix 
of the RHNs.
In this basis, the Dirac mass matrix in Eq. (\ref{eq:YD}) becomes
\begin{eqnarray}
Y_R\equiv Y_D V_{TB}P =\sqrt{2}\alpha
 \left(\begin{array}{ccc}
  0 & 0 & 0 \\
  0 & 0 & -1 \\
  0 & 0 & 1
 \end{array}\right)\label{eq:YR}.
\end{eqnarray}
Note that $\alpha$ can always be real by suitable redefinitions of 
the left-handed leptons.
As pointed out in Ref. \refcite{jarlskog}, in this basis, the RHN of 
the first family can be regarded as a non-interacting massless neutrino.
By omitting this field, we can move to $3\times2$ dimensional Dirac mass
matrix basis and rewrite Eq. (\ref{eq:YR}) as
\begin{eqnarray}
Y_R= \sqrt{2}\alpha
 \left(\begin{array}{cc}
  0 & 0 \\
  0 & -1 \\
  0 & 1
 \end{array}\right),
\end{eqnarray}
with $D_R=\diag(M_2, M_3)$.
%Then, t
The mass matrix of the light neutrinos is written as
\begin{eqnarray}
{\cal M}^{\nu}=v^2 Y_R D_R^{-1} Y_R^T
=\frac{2\alpha^2 v^2}{M_3}
\left(\begin{array}{ccc}
 0 & 0 & 0 \\
 0 & 1 & -1 \\
 0 & -1 & 1
\end{array}\right).
\end{eqnarray}
This matrix can be diagonalized with only one maximal angle and
has only one non-zero eigenvalue.
%where we choose $m_3$ as a non-zero one.
Hence, there are two interacting and one non-interacting 
massless neutrinos and  no CP violating phase in the MNS matrix.
Clearly, it is inconsistent with the experimental data of existing at least two
massive light neutrinos and large mixing angles.

In order to obtain a realistic model, we introduce symmetry breaking terms 
in Eq. (\ref{eq:YD}), given by
\begin{eqnarray}
Y_D=
 \left(\begin{array}{ccc}
  0 & 0 & 0 \\
  0 & \alpha & -\alpha \\
  0 & -\alpha & \alpha
 \end{array}\right)
 +
   \left(\begin{array}{ccc}
  \frac{1}{4}\Delta & \frac{1}{2}\Delta & 0\\
  \frac{1}{2}\Delta & \Delta & 0 \\
  0 & 0 & 0
 \end{array}\right) \label{eq:YD2}.
\end{eqnarray}
Note that the breaking terms violate both the permutation symmetry in 
Eq. (\ref{eq:twFL}) and the $Z_2$ symmetry, but still preserve the 
translational symmetry
so that the first family light neutrino remains massless.
Note also that
although we could introduce breaking terms for the Majorana mass matrix as well,
we only focus on the effect from the Dirac mass matrix in the following discussions.

In the diagonal basis of the RHNs, the Dirac mass matrix
can again be reduced to an $3\times 2$ dimensional matrix
%\footnote{This feature is ensured because the breaking terms still 
%preserve the translational symmetry.}
and becomes
\begin{eqnarray}
Y_R = 
 \frac{1}{2}
  \left(\begin{array}{cc}
  \frac{\sqrt{3}}{2}\Delta e^{i\phi_R /2} &
  -\frac{\sqrt{2}}{2}\Delta \\
  \sqrt{3}\Delta e^{i\phi_R /2} & 
  -2\sqrt{2}\alpha - \sqrt{2}\Delta \\
  0 & 2\sqrt{2}\alpha
 \end{array}\right).\label{eq:YR2}
\end{eqnarray}
In what follows, we consider the basis where $\alpha$ is real but 
$\Delta$ is complex, $\Delta \equiv |\Delta|e^{i\phi_{\Delta}}$.
The mass matrix of the light neutrinos is given by
\begin{eqnarray}
{\cal M}^{\nu}= \frac{v^2}{M_3}\left[2\alpha^2
 \left(\begin{array}{ccc}
 0 & 0 & 0 \\
 0 & 1 & -1 \\
 0 & -1 & 1
 \end{array}\right)
+\frac{\alpha\Delta}{2}
 \left(\begin{array}{ccc}
 0 & 1 & -1 \\
 1 & 4 & -2 \\
 -1 & -2 & 0
 \end{array}\right)
 +\frac{\Delta^{'2}}{8}
 \left(\begin{array}{ccc}
 1 & 2 & 0 \\
 2 & 4 & 0 \\
 0 & 0 & 0
 \end{array}\right)
\right],
%+{\cal O}(M_3/M_2),\hspace{0.5cm}
\label{eq:Mn}
\end{eqnarray}
where the second and third terms are responsible for the deviations from the 
tribimaximal mixing with
\begin{eqnarray}
\Delta^{'2}=|\Delta|^2 e^{2i\phi_\Delta}
   \left[1+\frac{3}{2}\frac{M_3}{M_2}e^{i\phi_R} \right] ,\label{eq:Delp}
\end{eqnarray}
while $\phi_\Delta$ and $\phi_R$ generate CP violation in the MNS matrix.
Here, we define the MNS matrix as 
\begin{eqnarray}
V_{MNS}=V_{TB}\ \delta V\ \Omega=\frac{1}{\sqrt{6}}
 \left(\begin{array}{ccc}
  2 & \sqrt{2} & 0 \\
  -1 & \sqrt{2} & -\sqrt{3} \\
  -1 & \sqrt{2} & \sqrt{3}
 \end{array}\right)
 \left(\begin{array}{ccc}
 1 & 0 & 0 \\
 0 & c_{\theta} & s_{\theta} e^{-i\delta} \\
 0 & -s_{\theta} e^{i\delta} & c_{\theta}
 \end{array}\right)\Omega,\label{eq:MNS1}
\end{eqnarray}
where $s_{\theta}=\sin\theta$ ($c_{\theta}=\cos\theta$) with
\begin{eqnarray}
\tan{2\theta}=
-\frac{\sqrt{6}(\alpha\Delta+\Delta^{'2} /4)e^{i\delta}}
{(4\alpha^2+2\alpha\Delta + \Delta^{'2}/4)e^{2i\delta}-3/8\Delta^{'2} }
\equiv 
-\frac{{\cal I}e^{i\delta}}{{\cal J}e^{2i\delta}-{\cal K}}\ ,
\label{eq:theta}
\end{eqnarray}
$\delta$ is a Dirac-type CP phase which has to satisfy
\begin{eqnarray}
\delta=
-\frac{i}{2}\ln
\left[\frac{{\cal IJ^{*}+I^{*}K}}{{\cal I^{*}J+IK^{*}}}\right]
\label{eq:dirac},
\end{eqnarray}
to guarantee the right hand side of Eq. (\ref{eq:theta}) to be real, 
and $\Omega=\diag(1, e^{i\gamma /2},1)$ is a diagonal Majorana-type 
CP phase matrix.
Note that relations between our definitions of the Dirac and Majorana phases 
and those in  Particle Data Group\cite{pdg} (PDG) are given in Ref. \refcite{FL-LG}.
The mixing angles are given by
\begin{eqnarray}
&&\sin^2\theta_{13}=\frac{1}{3}s^2_{\theta}, \label{eq:rct}\\
&&\sin^2\theta_{12} \simeq \frac{1}{3}(1-s^2_{\theta}), \label{eq:sol}\\
&&\sin^2\theta_{23} \simeq \frac{1}{2}-\frac{1}{6}s^2_{\theta}
    -\frac{\sqrt{6}}{3}s_{\theta}c_{\theta}\cos\delta. \label{eq:atm}
\end{eqnarray}
The mass matrix in Eq. (\ref{eq:Mn}) is diagonalized by Eq. (\ref{eq:MNS1}), 
leading to the masses of the light neutrinos to be
\begin{eqnarray}
&&m_1=0, \\
&&m_2 = \frac{v^2}{M_3}\left|4\alpha^2 s^2_{\theta} e^{2i\delta}
+\alpha\Delta(\sqrt{6}s_{\theta}c_{\theta} e^{i\delta}
+2s^2_{\theta}e^{2i\delta})
\right.\nonumber\\ &&\hspace{4cm}\left.
+\frac{\Delta^{'2}}{4} (\sqrt{6}s_{\theta}c_{\theta} e^{i\delta}
+s^2_{\theta}e^{2i\delta}+3/2 c^2_{\theta})
\right|
% \simeq \frac{4\alpha^2 v^2}{M_3} s^2_{\theta}
 , \label{eq:m2}\\
&&m_3 = \frac{v^2}{M_3}\left|4\alpha^2 c^2_{\theta}
+\alpha\Delta(-\sqrt{6}s_{\theta}c_{\theta}e^{-i\delta}+2c^2_{\theta})
\right.\nonumber\\ &&\hspace{4cm}\left.
+\frac{\Delta^{'2}}{4} (-\sqrt{6}s_{\theta}c_{\theta} e^{-i\delta}
+3/2 s^2_{\theta}e^{-2i\delta}+c^2_{\theta})
\right|
% \simeq \frac{4\alpha^2 v^2}{M_3} c^2_{\theta}
 . \label{eq:m3}
\end{eqnarray}
The Majorana phase is given by
 \begin{eqnarray}
 \gamma&=&-\gamma_2 + \gamma_3\,,
 \label{eq:gamma}
 \end{eqnarray}
  where
\begin{eqnarray}
\sin\gamma_2 = \frac{{\rm Im} [m_2]}{|m_2|} ,\ \ 
\sin\gamma_3 = \frac{{\rm Im} [m_3]}{|m_3|}. \label{eq:majo}
\end{eqnarray}
From Eqs. (\ref{eq:dirac}), (\ref{eq:m2}), (\ref{eq:m3}), (\ref{eq:gamma}) and (\ref{eq:majo}), 
one can see that the Dirac and Majorana phases are originated from
$\phi_R$ and $\phi_\Delta$.

\subsection{CP violation}
Our model possesses two CP violating phases: $\phi_R$ and $\phi_\Delta$,
plus four real parameters: $\alpha$, $|\Delta|$, $|2A+B|$ and $|B|$.
These six theoretical parameters can be fixed by six physical quantities.
In our calculations, we will use the following best-fit values with 
$1\sigma$ errors:
\begin{eqnarray}
 &&\Delta m_{21}^2=(7.65_{-0.20}^{+0.23})\times 10^{-5}\ \ev^2,\ \ 
 |\Delta m_{31}^2|=(2.40_{-0.11}^{+0.12})\times 10^{-3}\ \ev^2,\label{eq:Dm}\\
 &&\sin^2\theta_{12}=0.304_{-0.016}^{+0.022},\ \ 
 \sin^2\theta_{23}=0.50_{-0.06}^{+0.07},\label{eq:mix}
\end{eqnarray}
from Ref. \refcite{osi} and
\begin{eqnarray}
 \sin^2\theta_{13}=0.02\pm0.01 \label{eq:gbal}
\end{eqnarray}
from the global analysis in Ref. \refcite{global}, and take
\begin{eqnarray}
 M_3=8.0\times 10^{10}\ \gev,\hspace{0.3cm} M_3/M_2 = 5
\end{eqnarray}
as input parameters.
As we will discuss later, the masses of RHNs are determined to account 
for the measured value of the BAU.
%The remaining two quantities are the masses of the heavy neutrinos, $M_2$ and $M_3$.
%As we will discuss later, in order to account for the measured value of the BAU,
%they should be ${\cal O}(10^{10\sim11})\ \gev$, 
%corresponding to $\alpha \sim 0.004$ and $|\Delta| \sim 0.002$.
%If $M_2\gg M_3$, $\phi_R$ will be decoupled from low energy observables such as 
%light neutrino masses, mixing angles and CP violating phases in the MNS matrix.
%Hence, the leptogenesis ends up depending on phases 
%which cannot be observed by low energy experiments.
%On the other hand, in order to fit Eq. (\ref{eq:Dm}), $M_3$ cannot be 
%much larger than $M_2$ and their ratio is constrained by $M_3/M_2 \leq 6$.
%To illustrate our results, we take $M_3=8.0\times 10^{10}\ \gev$ and 
%$M_3/M_2 = 5$.

\begin{figure}[tb]
\begin{center}
\includegraphics*[width=0.6\textwidth]{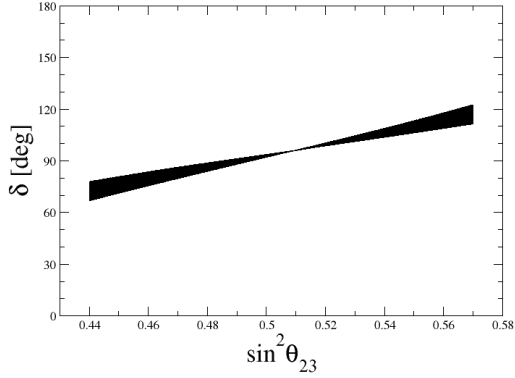}
\vspace*{8pt}
\caption{\label{fig:a-d}
The Dirac phase $\delta$ as a function of $\sin^2\theta_{23}$
with $M_3=8.0\times 10^{10}\ \gev$ and $M_3/M_2 = 5$.}
\end{center}
\end{figure}

%From Eqs. (\ref{eq:rct}) and (\ref{eq:atm}), one can see that
%the Dirac phase $\delta$ can be described by $\sin\theta_{23}$ and 
%$\sin\theta_{13}$.
%Furthermore, from Eqs. (\ref{eq:sol}) and (\ref{eq:mix}), one obtains that 
%\begin{eqnarray}
%0.0073<\sin^2\theta_{13}\,.%<0.0453
%\label{eq:theta13}
%\end{eqnarray}
By using Eqs. (\ref{eq:rct}) - (\ref{eq:atm}) and $1\sigma$ values of 
$\sin^2\theta_{23}$ and $\sin^2\theta_{13}$, we can estimate the range of 
$\delta$ to be
\begin{eqnarray}
67^\circ<\delta<122^\circ\,.
\label{eq:delta}
\end{eqnarray}
In Fig. \ref{fig:a-d}, we show the numerical result of $\delta$ as a function 
of $\sin^2\theta_{23}$.
As can be seen from the figure, the result is coincident with 
Eq. (\ref{eq:delta}) very well.

\begin{figure}[t]
\begin{center}
\includegraphics*[width=0.6\textwidth]{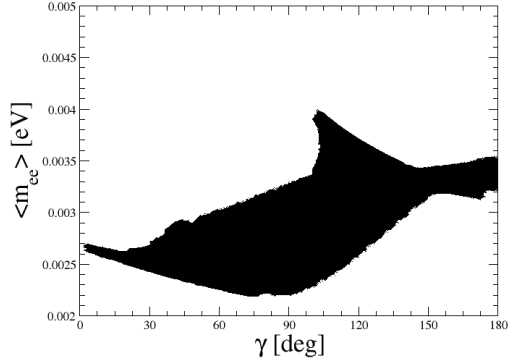}
\vspace*{8pt}
\caption{\label{fig:0nbb}
The effective mass $<m_{ee}>$ as a function $\gamma$ 
with $M_3=8.0\times 10^{10}\ \gev$ and $M_3/M_2 = 5$.}
\end{center}
\end{figure}

In contrast, $\gamma$ has a wide allowed range and it
could have an impact on the neutrinoless double $\beta$ decay due to the 
effective Majorana mass 
\begin{eqnarray}
<m_{ee}>=\left| \sum_{i=1}^3 m_i (V_{MNS})_{1i}^2 \right|
\end{eqnarray}
since the Dirac phase as well as the individual neutrino mass can be determined 
within some ranges in our model.
In Fig. \ref{fig:0nbb}, we give the effective mass as a function of $\gamma$.
Unfortunately, the  predicted values of $<m_{ee}>$ in our model are around
$(2.2-4.1)\times 10^{-3}$~eV, which are too small to be detected 
in the current and upcoming experiments.
For instance, the order of the present sensitivity at the CUORICINO experiment is 
$10^{-1}\ \ev$, while that of the proposed CUORE detector is 
$10^{-2}\ \ev$.\cite{0nbb}
Nevertheless, we would like to emphasize that more dedicated experiments 
in future are needed in order to determine  the Majorana phase.

Finally, we would like to briefly remark on the possibility to test 
our model.
As our model predicts the following novel relation
\begin{eqnarray}
\sin^2\theta_{13}\simeq 1/3-\sin^2\theta_{12}\label{eq:s1312}
\end{eqnarray}
based on Eqs. (\ref{eq:rct})
and (\ref{eq:sol}), more precise determinations of mixing angles would 
provide us a chance to rule out or confirm the model in future.
For instance, the smaller value of $\sin^2\theta_{12}$, which is 
$\sin^2\theta_{12}=0.304^{+0.000}_{-0.016}$, results in 
$\sin^2\theta_{13}=0.0293\sim 0.0453$ which goes beyond the $1\sigma$ 
ranges given in Ref. \refcite{osi} and the global analysis in Ref. \refcite{global}.
On the other hand, the larger value of $\sin^2\theta_{12}=0.304^{+0.022}_{-0.000}$
corresponding to $\sin^2\theta_{13}=0.0073\sim 0.0293$ is well coincident 
with Refs. \refcite{osi} and \refcite{global}.

\section{Leptogenesis}
As discussed in the previous section, our model results in non-zero values of 
$\delta$ and $\sin\theta_{13}$ as shown in 
Eq. (\ref{eq:delta}) and Eq. (\ref{eq:s1312}) with Eq. (\ref{eq:mix}), 
respectively.
This means that the CP symmetry is always violated in the lepton sector 
even if there is no Majorana phase $\gamma$. 
In this section, we consider the unflavored leptogenesis mechanism\footnote{
The importance of the flavor effects is discussed in 
Refs. \refcite{flavor1,flavor2} and \refcite{flavor3}.}
via the out-of-equilibrium decays of the heavy RHNs.
The CP violating parameter in the leptogenesis 
 due to the i-th heavy RHN decays is written as
\begin{eqnarray}
\varepsilon_i =-\frac{1}{8\pi}\sum_{j\neq i}
\frac{{\rm Im}[(Y_R^{\dag} Y_R)_{ji}^2]}{(Y_R^{\dag} Y_R)_{ii}}
\ F\left(\frac{M_j^2}{M_i^2}\right), \label{eq:gCP}
\end{eqnarray}
where $i,j=2$ or $3$, $F(x)$ is given by
\begin{eqnarray}
F(x)=\sqrt{x}\left[\frac{1}{1-x}+1-(1+x)\ln\frac{1+x}{x} \right],
\end{eqnarray}
$Y_R$ is the Dirac mass matrix in the diagonal basis of the right-handed 
neutrinos and charged leptons, given in Eq. (\ref{eq:YR2}), 
with the first (second) column  referred as $Y_{R_{j2}}$ ($Y_{R_{j3}}$).
The dilution factor $\kappa_i$ is approximately given by\cite{earlyU}
\begin{eqnarray}
 \kappa_i\simeq \frac{0.3}{r_i (\ln r_i )^{0.6}}\ ,
\end{eqnarray}
where
\begin{eqnarray}
r_i =
\frac{\Gamma_i}{H|_{T=M_i}} = \frac{M_{pl}}{1.66\sqrt{g_*}M_i^2}
\frac{(Y_R^{\dag} Y_R)_{ii}}{16\pi}M_i
\end{eqnarray}
with $M_{pl}=1.22\times 10^{19}\ \gev$ and $g_{*}=106.75$.
The net BAU is found to be
\begin{eqnarray}
\eta_B = \frac{n_B}{n_{\gamma}} = 7.04\frac{\omega}{\omega-1}
\frac{\kappa_2 \varepsilon_2 + \kappa_3 \varepsilon_3}{g_{*}},
\end{eqnarray}
where $\omega=28/79$.
\begin{figure}[t]
\begin{center}
\includegraphics*[width=0.6\textwidth]{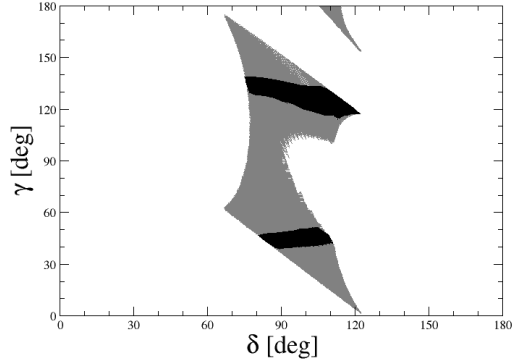}
\vspace*{8pt}
\caption{
Allowed regions in $\delta-\gamma$ plane
with $M_3=8.0\times 10^{10}\ \gev$
and $M_3/M_2 = 5$, 
where the gray and black regions correspond to
those fitted by only the neutrino oscillation and with
WMAP data at $1\sigma$, respectively.
 }\label{fig:d-m}
\end{center}
\end{figure}
Here, instead of showing some complex analytic calculations, we only give 
the numerical results.
In Fig. \ref{fig:d-m}, 
we show the allowed regions in $\delta-\gamma$ plane
with $M_3=8.0\times 10^{10}\ \gev$
and $M_3/M_2 = 5$, 
where the gray and black regions represent to those fitted by only the 
neutrino oscillation and with $1\sigma$ WMAP\cite{wmap} bound 
$\eta_B = (6.1_{-0.2}^{+0.2})\times 10^{10}$, respectively.
One can easily see that there is an explicit connection between the leptogenesis 
and the phases in the MNS matrix.
Especially, the Majorana phase is closely related to the leptogenesis 
and limited to two narrow regions. 

\section{Conclusion}
We have considered the twisted FL symmetry which can successfully generate 
the TBM neutrino mixing.
%And then, w
We have applied the symmetry to the seesaw framework and shown 
a specific model which is well consistent with current neutrino oscillation data.
%Next, w
We have studied the BAU through the leptogenesis mechanism
in the model and
%We have 
found that the net baryon asymmetry is directly connected with the 
CP violating Dirac and Majorana phases in the MNS matrix.
% and.
%In particular, the Majorana phase has been limited to the narrow regions by 
%the WMAP constraint.
\\

\section*{Acknowledgments}
This work is supported in part by the National Science Council of ROC under 
Grant \#s: NSC-95-2112-M-007-059-MY3 and NSC-98-2112-M-007-008-MY3
and by the Boost Program of NTHU.

\end{document}